\documentclass[12pt,preprint,epsf]{aastex}

\def\spose#1{\hbox to 0pt{#1\hss}} 
\def\simlt{\mathrel{\spose{\lower 3pt\hbox{$\mathchar"218$}}      
\raise 2.0pt\hbox{$\mathchar"13C$}}} 
\def\simgt{\mathrel{\spose{\lower 3pt\hbox{$\mathchar"218$}}      
\raise 2.0pt\hbox{$\mathchar"13E$}}} 
\def\eg{{\rm e.g.~}}  
\def\etal{{\rm et~al.~}} 

\received{May 10, 2012}
\revised{July 11, 2012} 
\articleid{LET27412R}{} 

\lefthead{Mould} 
\righthead{``Dis''integrated light} 

\begin{document}

\title{Galaxy evolution from ``dis''integrated light}

\author{Jeremy Mould} 
\affil{Centre for Astrophysics and Supercomputing, Swinburne University} 
\authoraddr{E-mail: jmould@swin.edu.au} 


\begin{abstract}
Masking the horizontal branch and giant stars allows unambiguous measurements of mean age and metallicity in simple old stellar populations from metal and hydrogen line strengths. Billion year resolution is possible in the luminous halos of early type galaxies, constraining formation models. Most of the nuisance parameters in stellar evolution are avoided by isolating the main sequence for analysis. The initial mass function and $s$-process element diagnostics may also be accessible. Integral field spectrographs have an significant advantage for this work, which is confusion limited by the presence of bright stars in medium to high surface brightness applications.
\end{abstract}

\section{Introduction}
We know the age of the universe from the expansion rate and from the nuclear burning rate of the oldest stars. The agreement between these two approaches is one of the pillars of the standard cosmology (Mould 1999). To learn about galaxy formation we must be able to age date the halos of galaxies. This is difficult because a notorious degeneracy between age and metallicity affects the integrated spectra of galaxies (Faber 1973, Mould 1978, Schmidt, Bica \& Dottori 1989). In this Letter we show that it is possible to deal with this problem by concentrating on the main sequence turnoff, where the age information in stellar populations resides, and by seeking separate spectral diagnostics of turnoff temperature and metallicity. Age resolution of a billion years is possible. Given the level of testing that main sequence evolution and the standard solar model have been subjected to, one can be reasonably confident about the uncertainties. The nuisance parameters of post main sequence stellar evolution: the mixing length to pressure scale height ratio, the Reimers mass loss parameter, the convective overshoot parameter and the horizontal branch (HB) color distribution are avoided. 

A comprehensive study of the spectra of old stellar populations has been made by Worthey (1994).
Mendel \etal 2007 point out that modelling of HB morphologies is difficult, as the interplay of contributing effects is not known well enough to be modeled purely from theory. Prescriptive methods have tended to be adopted. Maraston \& Thomas (2000) find that a mass-loss prescription is able to reproduce the strong Balmer lines found in old elliptical galaxy populations, as well as the trends of increasing H$\beta$ line strengths in low-metallicity Galactic globular clusters.
Methods of breaking the HB/age degeneracy in the Balmer lines have been discussed by Schiavon \etal (2004).

\section{Isolating the main sequence}
Many of the galaxies we wish to study in a broad analysis of the formation of the first stars across the Hubble sequence of galaxies are at the distance of the Virgo cluster, where the main sequence turnoff of old stellar populations is at magnitude 35, intractable for current and planned telescopes. However, if we place our 30 meter class telescope’s spectrograph slit across the outer parts of NGC 4472, say, and the seeing is good, we can in long exposures recognise individual horizontal branch stars at magnitude 32.  If we study the light between these stars and further exclude stars with red colors (subgiants), we have a sample of turnoff stars, which are ideal for age and metallicity study.
A quantitative example would be a study of a stellar population at 16 effective radii, 8.3 mag down from the central surface brightness of a de Vaucouleurs profile. If the local surface brightness is 27.5 mag/s$^2$, a square arcsec would contain 2000 L$_\odot$ at Virgo. This would be composed (by the fuel burning theorem, Greggio \& Renzini 2011, Renzini \& Buzzoni 1986) of some 500 main sequence stars and 4 horizontal branch stars\footnote{A referee has pointed out that a similar number of red giants would also be present and that analysis of their individual spectra would be valuable. Indeed this would be worth pursuing in order to measure detailed chemical composition. But it would not constrain age so well, which is our present emphasis.}. This is a feasible proposition for a large telescope in natural seeing. Sky subtraction is a challenge at this surface brightness, of course. Working closer to the nucleus would require adaptive optics.

An integral field unit spectrograph has a great advantage in this application. If the fibers are critically matched to the seeing, 4 fibers are lost in order to mask each horizontal branch star. However, in practise, rather than waste the light of HB stars, one would simply separate their spectra and make
use of this information to constrain other parameters. The same goes for subgiant stars.

\section{Separating age and metallicity}
Consider two diagnostics, H$\beta$ and the Ca K line, readily measurable in blue metal poor spectra. We write their line strengths, equivalent widths for example:
$$\beta = \beta(g, T, Z)$$
$$CaK = CaK(g, T, Z) \eqno(1)$$
in the usual notation with g = log(gravity), T = log(effective temperature), and Z = log(metal fraction).
Expanding around the main sequence turnoff we have
$$\delta \beta = \frac{\partial \beta}  {\partial g} \delta g + \frac{\partial \beta} {\partial T} \delta T + \frac{\partial \beta}  {\partial Z} \delta Z$$ 
$$\delta CaK = \frac{\partial CaK}  {\partial g} \delta g + \frac{\partial CaK}  {\partial T} \delta T + \frac{\partial CaK}  {\partial Z} \delta Z \eqno(2)$$ 
We also have for turnoff stars
$$\delta g = \frac{\partial g}  {\partial t} \delta t + \frac{\partial g} {\partial Z} \delta Z$$
$$\delta T = \frac{\partial T} {\partial t} \delta t + \frac{\partial T} {\partial Z} \delta Z \eqno(3)$$
where t is the log of the age of the simple stellar population. To evaluate these age and metallicity dependences, we use Padova YZVAR isochrones for helium fraction Y = 0.23 (Bertelli \etal 2008). We find for Z = --4 that $\partial g / \partial t = \partial T / \partial t$ = --0.0062 at the oldest ages and $\partial g / \partial Z$ = 0.0945 \& $\partial T / \partial Z$ = --0.017.
Synthetic line strength calculations from stellar atmospheres furnish the remaining information we need. According to Kurucz models\footnote{http://wwwuser.oat.ts.astro.it/castelli/grids/gridm20k2odfnew/am20k2tab.html}(Figure 1)
at [Fe/H] = --2, $\partial \beta / \partial T$ = 1.1, while at 6000K $\partial \beta / \partial g$ = --0.08 and $\partial \beta / \partial Z$ = 0.03. 
Similar results are obtained by Onehag et al (2008).  
We find from the synthetic spectrum calculations using MOOG (Sneden \etal 2012) and current ATLAS models (see Kurucz 1970, 1986 for the heritage): 
$\partial CaK / \partial T$ = 50; $\partial CaK / \partial g$ = 0.5 and $\partial CaK / \partial Z$ = 3. Ca K is measured as an equivalent width (EW) in \AA.
MOOG calculates EWs by direct integration of the (continuum, line profile) difference. The EWs are already tabulated in the H$\beta$ case. Based on the solar spectrum, one might question how well defined the CaK continuum is. However, there is no issue in the galaxy halo case Z--Z$_\odot~<$ --1.
Solar abundance ratios are assumed and 2 km/s of microturbulence.
Strictly one should also broaden the stellar line profiles with a galaxy velocity distribution. However, EWs are invariant to such post-broadening, \eg for stellar rotation, instrumental broadening, or macroturbulence. Stellar population broadening has therefore not been done here.
Similar results to Figure 2 were obtained by Franchini \etal (2011). 
Thus we end up with
$$\delta \beta = -0.08(-0.124 \delta t + 0.0945 \delta Z) + 1.1(-0.124 \delta t – 0.017 \delta Z) + 0.03 \delta Z = -0.126 \delta t + 0.004 \delta Z$$
$$\delta CaK = 6.1 \delta t + 3.1 \delta Z \eqno(4)$$
One can formally eliminate metallicity or simply write: $\delta t ~\approx~ -8\delta \beta$.     

Now imagine comparing a 13 Gyr simple stellar population (main sequence only) with a 13 $\times$ 2.303$\delta t$  Gyr younger one. Subtracting the two spectra, one would see a $\delta \beta$
given approximately by equation (4). Stars on the main sequence of lower mass than the turnoff would tend to cancel out.

\section{Secondary parameters}
In the preceding section the assumption has been that a stellar spectrum depends on (g, T, Z) and that age can be deduced from hydrogen and metal lines to the limits of observational error.
However, the study of $\omega$ Cen has taught us that helium abundance is important and the systematics of [O/Fe] as a function of Z (\eg Pilachowski \& Armandroff 1996, Cohen \& Bell 1986) remind us that oxygen is the most abundant 
metal. Blue horizontal branch stars (Schiavon 2007) can tell us both helium abundance and oxygen abundance. In red giants the 6300\AA~ line measures oxygen abundance. Fibers that intercept the light from these stars can
therefore be used to diagnose oxygen and helium anomalies in disintegrated spectra. However, this is beyond the scope of the present Letter. For the time being it is reasonable to assume
that Z predicts O/Fe and that helium has its cosmological abundance in protogalactic halos.

Alpha element enhancement has also been a subject of interest in the
integrated light of early type galaxies. Figure 4 shows the prediction
for the Mg/Fe ratio of metal poor turnoff stars. The iron line is 100--200 m\AA~ in equivalent width.
Again, as a starting point in this Letter, it is assumed that $\alpha$/Fe is a constant in protogalactic halos and can be omitted from equation (1).
Figure 4 is a precursor of a more detailed study, which, fortunately, is made easier by the existence of appropriate Kurucz models.
The work of Dotter \etal (2007) notwithstanding, if $\alpha$/Fe can be omitted from equation (1), it can also be omitted from the age error budget, provided a non-$\alpha$ element metal line is also observed.

\section{Summary}
An observational error of 0.005 in $\beta$ 
corresponds to an uncertainty of 1 Gyr at 12 Gyr. This is useful for distinguishing chemodynamical models of rapid initial collapse from those of assembly through mergers up to redshift 2.
A full evolutionary synthesis model, following Leitherer \etal (1999) and Maraston \etal (2003), is now warranted to include light from the full main sequence, rather than just the turnoff itself. Other lines should also be considered, such as the Ca triplet and the Paschen lines in the near infrared. Foster \etal (2009, 2010, 2011) have pioneered the study of the Ca triplet in halos and globular clusters.

From the very beginning, studies of integrated light have sought to constrain the initial mass function (IMF) in early type galaxies (Spinrad \& Taylor 1971). Disintegrating the light will strengthen the observational approach to this problem too. Indices such as the Wing-Ford band (Wing \& Ford 1969, Wing, Cohen \& Brault 1977, van Dokkum \& Conroy 2011) will be more sensitive to IMF, when only main sequence light is analysed.

Barium lines are an indicator of $s$-process activity (secondary nucleosynthesis) in the earliest
stars. Ba II 4554\AA~ has 77 m\AA~ equivalent width at (T$_{eff}$, g, [Fe/H]) = (6000, 4, --2). The ratio of predicted equivalent widths at other effective temperatures\footnote{A referee has suggested that the conventional approach to measuring Ba/Fe in red giants may be superior. Indeed, Ba II lines are stronger on the giant branch, and if the giants can be isolated, they should be analyzed.}
is shown in Figure 3.
With a 100 km/s velocity dispersion this will be a 5\% depression in the continuum
of a metal poor main sequence spectrum. High signal-to-noise disaggregated\footnote{Possibly a better name than disintegrated.} spectra will be required.
Figures 3 and 4 show that $s$-process and $\alpha$ elements are accessible to the technique described here and the temperature dependence of these
line ratios is modest in the region which contributes most of the light.

\section*{References}
\noindent Bertelli, G. \etal 2008, A\&A, 484, 515\\
Cohen, J. \& Bell, R. 1986, ApJ, 305, 698\\
Dotter, A. \etal 2007, ApJ, 666, 403\\
Faber, S. 1973, ApJ, 179, 731\\
Foster, C. \etal 2011, MNRAS, 415, 3393\\
Foster, C. Forbes, D. Proctor, R. Strader, J., Brodie, J., \& Spitler, L. 2010, AJ, 139, 1566\\	
Foster, C. \etal 2009, MNRAS, 400, 2135\\
Franchini, M. \etal 2011, ApJ, 730, 117\\
Greggio, L. \& Renzini, A. 2011, Stellar Populations, Wiley Series in Cosmology\\
Kurucz, R. 1970, SAO Special Report 309\\
Kurucz, R. 1986, in Highlights of astronomy, v7 - Proc. 19th IAU General Assembly, Dordrecht: Reidel, p. 827-831.\\
Leitherer, C. \etal 1999, ApJS, 123, 3\\
Maraston C. \& Thomas D., 2000, ApJ, 541, 126\\
Maraston, C. \etal 2003, A\&A, 400, 823\\
Mendel, J., Proctor, R. \& Forbes, D. 2007, MNRAS, 379, 1618\\
Mould, J. 1978, ApJ, 220, 434\\
Mould, J. 1999, Bull AAS, 194, 3911\\
Onehag, A. \etal 2009, A\&A, 498, 527\\
Pilachowski, C. \& Armandroff, T. 1996, AJ, 111, 1175\\
Renzini, A. \& Buzzoni, A. 1986, in Spectral evolution of galaxies; Proc. 4th Workshop, Erice, Italy, Dordrecht: Reidel, p. 195-231\\
Schiavon, R. 2007, ApJS, 171, 146\\
Schiavon, R., Rose, J., Courteau, S. \& MacArthur, L. 2004, ApJ, 608, L33\\	
Schmidt, A., Bica, E., \& Dottori, H. 1989, MNRAS, 238, 925\\
Sneden, C., Bean, J. Ivans, I. Lucatello, S. \& Sobeck, J. 2012, Astrophysics Source Code Library, record ascl:1202.009\\
Spinrad, H. \& Taylor, B. 1971, ApJS, 22, 445\\ 
van Dokkum, P. \& Conroy, C. 2011, ApJ, 735, L13\\
Wing, R., Cohen, J. \& Brault, J. 1977, ApJ, 216, 659\\
Wing, R. \& Ford, WK 1969, PASP, 81, 527\\
Worthey, G. 1994, ApJS, 95, 107 

\acknowledgements
\noindent I acknowledge helpful discussions with Chiara Tonini.
I am grateful to Robert Kurucz for the use of his H$\beta$ calculations\footnote{
http://kurucz.harvard.edu/grids.html} . And to Chris Sneden for use of the MOOG spectrum synthesis program.
Thanks also to Duncan Forbes for reading a draft and to an anonymous referee
for sound advice.

\pagebreak

\begin{figure}[h]
\begin{center}
\includegraphics[angle=-90, width=\textwidth]{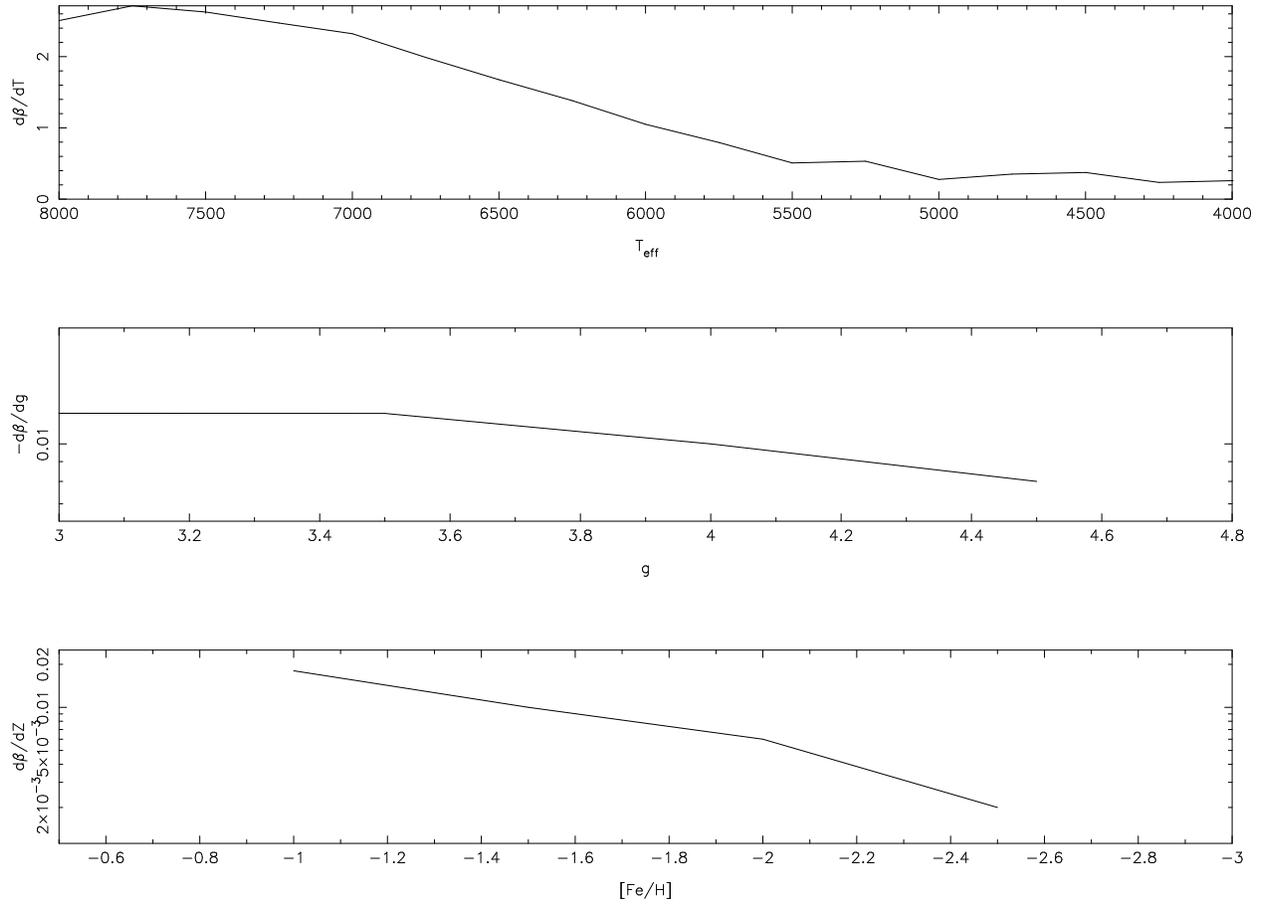}
\end{center}
\caption{ Dependence of H$\beta$ on temperature, gravity, and metallicity.
The reference point for the top plot is g = 4, [Fe/H] = --2, for the center
plot is T$_{eff}$ = 6000K, [Fe/H] = --2, and for the bottom plot is 
T$_{eff}$ = 6000K, g = 4.}
\end{figure}

\begin{figure}[h]
\begin{center}
\includegraphics[angle=-90, width=\textwidth]{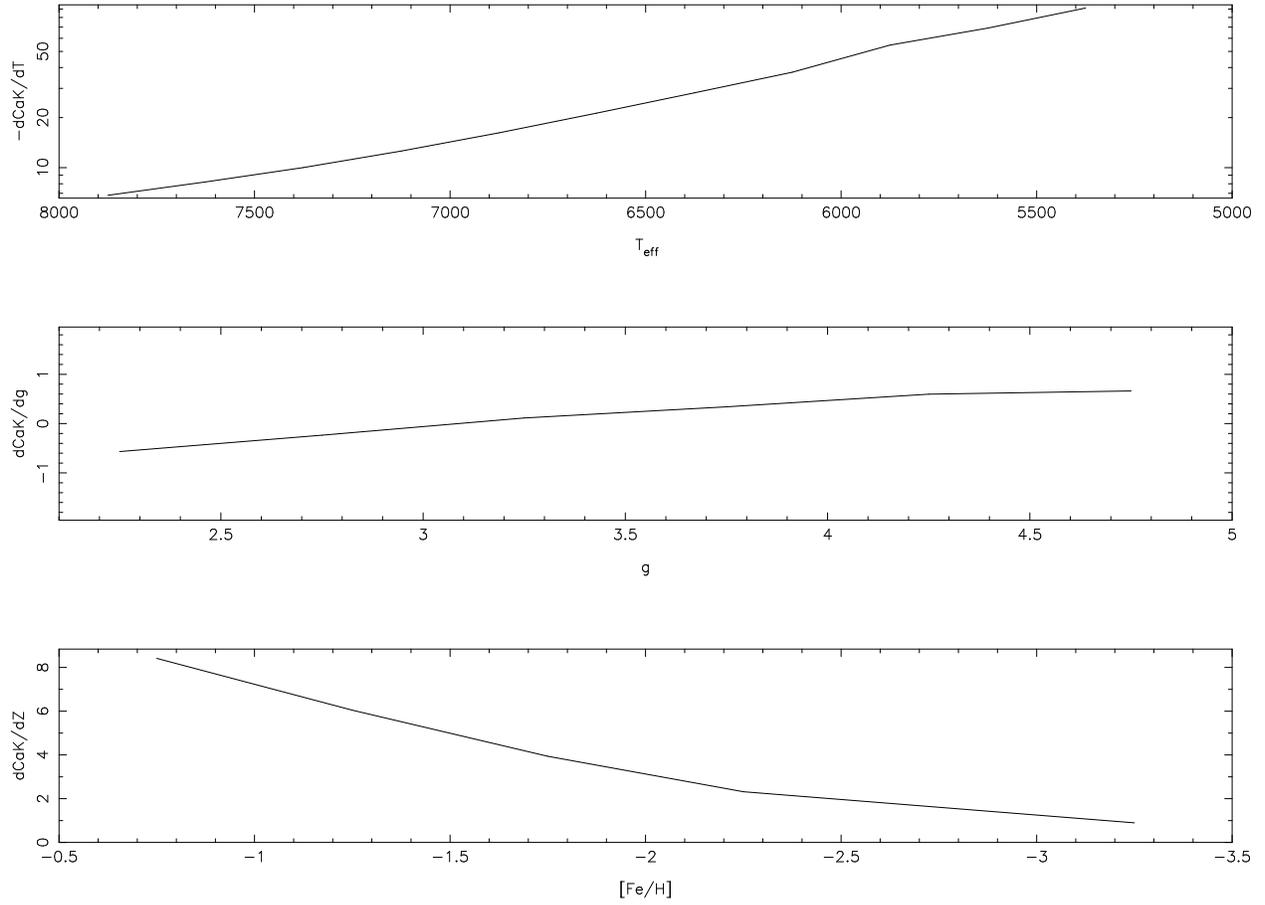}
\end{center}
\caption{ Dependence of Ca K equivalent width on temperature, gravity, and metallicity.
The reference points are the same as for Figure 1.}
\end{figure}
\begin{figure}[h]
\begin{center}
\includegraphics[angle=-90, width=0.5\textwidth]{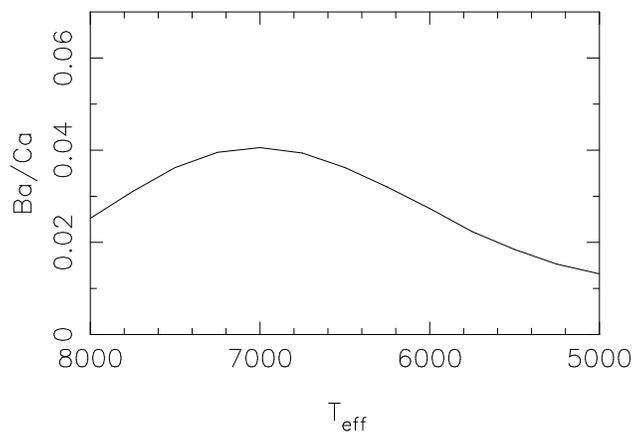}
\end{center}
\caption{ Dependence of the ratio of Ba 4554 to Ca K equivalent width on temperature.
The reference point is g = 4 and [Fe/H] = --2. This curve reflects the behavior of two chemically
similar elements in the same ionization state and ground state excitation, separated on the
curve of growth by 4 orders of magnitude in abundance.}
\end{figure}
\begin{figure}[h]
\begin{center}
\includegraphics[angle=-90, width=0.5\textwidth]{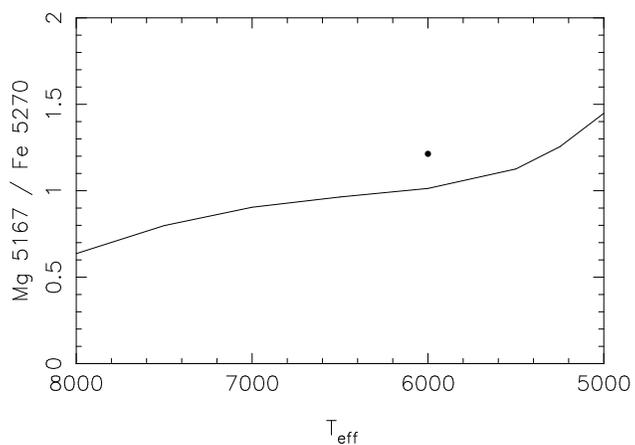}
\end{center}
\caption{ Dependence of the ratio of the equivalent width of the first line of the Mg triplet to Fe 5269.54\AA~ on temperature.
The reference point is the same as for Figure 3. The single point indicates
an alpha element enhancement of 0.4 dex.}
\end{figure}

\end{document}